\documentclass[aps,preprint,12pt]{revtex4}
\usepackage{epsf}
\begin{document}
\columnsep     38pt
\topmargin    -30pt
\oddsidemargin  5pt
\parsep  3pt plus 1pt minus 1pt

\title{ Rescattering and finite formation time effects  in inclusive
and exclusive
electro-disintegration of nuclei within a relativistic approach:
1. The deuteron}
\author{M.A.Braun} \altaffiliation{Permanent address:
 Dept.  of High-Energy Phys., S. Petersburg Univ., 198904 S. Petersburg,
 Russia}
\author{C.Ciofi degli Atti}
\author{L.P.  Kaptari}
\altaffiliation{Permanent address: Bogoliubov Lab. Theor. Phys.
JINR, Dubna,  Russia}
\affiliation{
 INFN, Sezione di Perugia, via A. Pascoli, Perugia, I-06100, Italy}
\date{\today}

\def\beq{\begin{equation}}
\def\eeq{\end{equation}}
\def\noi{\noindent}
\def\bk{{\bf k}}
\def\bk'{{\bf k}'}
\def\tka{\tilde{k}_1}
\def\tkb{\tilde{k}_2}
\def\tr{\tilde{r}}

\medskip
\vspace{1 cm}

\begin{abstract}
 The rescattering contribution to the inclusive and exclusive
deuteron electro-disintegration at the values of the Bjorken scaling variable $x=1$, as well as in
 the so called {\it cumulative} region ($x>1$) is calculated within a
  relativistic
 approach based on the Feynman diagram formalism taking also into
 account colour transparency
 effects by  the inclusion of the  {\it finite formation time} (FFT) of
 the ejected nucleon
 via the introduction of the dependence of the scattering amplitude of
 the ejectile upon its
 virtuality. In the cumulative region  the FFT effects
 which
  result  from the real part of the ejectile  propagator are taken into account.
  It is found that the relative weight
of the rescattering  steadily grows with $x$
becoming  of the order of unity at $x>1.4\div 1.5$.
At such values of $x$  the  finite formation
time effects become fairly visible, which may serve for
their study at relatively small value of the four-momentum transfer
  $Q^2$. The relativistic rescattering contribution is  compared with
  the Glauber rescattering, which is shown to be not valid  in the cumulative region
  starting from  $x>1.4$.
\end{abstract}

\maketitle
\newpage
\section{Introduction}

Quasi-elastic inclusive and exclusive  high energy electron scattering off
nuclei of the
type $A(e,e')X$, $A(e,e'h)X$, etc, where $h$ denotes a hadron,
represent powerful tools to investigate both the properties of hadronic
matter, as well
as basic QCD predictions,  provided one is able to evaluate the effects
of the
Final State Interaction (FSI) of the lepto-produced hadrons.
 As a matter of fact, the possibilities to get information on basic
 properties
of bound hadrons, such as, for example  their  momentum and energy
distributions, crucially
depend upon the ability to estimate to which extent
FSI effects  destroy the direct link between the measured  cross section
and the hadronic
 properties before interaction with the probe, so the main aim is to minimize
 the effects of FSI.
 At the same time, the necessity for an accurate treatment of  FSI,
 stems from the QCD prediction  that at large $Q^2$ FSI should vanish
because of  Color Transparency (CT)
\cite{zamo}- \cite{brodsky}
 (for  recent reviews on the subject, see e.g. \cite{color},\cite{niko}),
 according to which
  the  rescattering
 amplitudes of the hit nucleon (the {\it ejectile})  corresponding
 to various excited states
 interfere destructively. This is equivalent to the Gribov inelastic
 corrections \cite{gribov} to the
Glauber multiple scattering approximation. Thus the
 experimental investigation of CT is thought to be the detection
  of possible differences between  experimental data
 and predictions of standard Glauber multiple scattering calculations of FSI.
 While there is a vast consensus on the qualitative features of the effect,
 the quantitative
 aspects are still under debate and many details in the theoretical
 formulation of the problem
 are not unique. Most problems arise for the following reason:
 whereas the Glauber approach  is a workable and reliable approximation
 to the scattering of
 hadrons off nuclei, problems may arise when one is treating the
 rescattering
 of a hadron produced in the medium. Therefore  one faces both the
 problem of calculating
 the usual rescattering of the produced hadron and the introduction of
 CT effects.
  The standard approach
 to CT is formulated in terms of a multi-channel problem, the vanishing of
 FSI
 taking place via  the cancellation between
contributions from  channels with various  excitations of the nucleon.
This cancellation is expected to become
fully operative
in a high energy regime, when, at least, several
nucleon states are excited.
Where
precisely such a regime is located
is not  yet quite clear. Present evidence from experimental \cite{exp} and
theoretical  \cite{comparison} investigations  of quasi-elastic
$A(e,e'p)X$  reactions    seems to indicate that CT effects
are not visible up to $Q^2$ of the order 20 (GeV/c)$^2$. The treatment of
CT within the multi-channel Glauber approach
inevitably involves many poorly known quantities, related both
to the structure of the
tower of excited nucleon states and to diagonal
and non-diagonal amplitudes for their rescattering.
It  also  usually leads to rather heavy expressions for
numerical evaluation. This makes the
predictions of various approaches not unique and sometimes even conflicting.
 A formally different, though physically equivalent approach for the
 treatment
of  FSI in lepto-production processes has been recently developed
\cite{BCT}, based upon a  direct
evaluation of Feynman diagrams and assuming an explicit dependence
of the nucleon-nucleon scattering amplitude on the virtuality of the
ejectile;
by this way one introduces a finite formation time (FFT) for the produced
hadronic
state to become a physical hadron. In this approach,
which, for a single rescattering is
equivalent to the CT in a two-channel approach,
 FSI vanish
because with increasing $Q^2$ the virtuality also increases and
the amplitudes diminish. The approach has been applied to the inclusive
process
$A(e,e')X$ at the value of the Bjorken scaling variable
$x=Q^2/(2M\nu) \simeq 1$.
 In this paper we generalize the approach to  $x>1$ and to exclusive
 processes $A(e,e'p)B$ for which a large wealth of papers, aimed at
 investigating
 CT effects, already  exists at $x=1$,  where the energy in the final
state is sufficiently high to produce physical nucleon resonant states.
The aim of this paper is to investigate CT (FFT) effects also  at
$x > 1$, i.e.
in the so called {\it cumulative} region. We will present here
 the results for the deuteron, leaving a discussion for complex nuclei
 for  subsequent paper.
%In fact the CT or FFT influence in collisions with nuclei can
%hardly be treated theoretically due to formidable calculational
%problems. A much easier object is the deuteron. Of course the
%final-state interactions (FSI), sensitive to CT or FFT, are generally
%small in the deuteron due to its relatively large size. However they
%can be predicted with greater reliability than for heavy
%nuclei and with growing precision of experimental studies can
%well be checked.
 The cumulative region  has not been much investigated
 in the past  on the
basis of the
 conjecture that the influence of CT  should
be much smaller than at $x \simeq 1$, since
the available final state energy   may be insufficient to
produce those excited states of the hit nucleon,
 whose contribution is to cancel the one coming from
the lowest scattering states. Nonetheless the importance of the investigation of CT effects
 in the cumulative region has been stressed  Ref.  \cite{kope},
 where it has been shown that,
 due to the different role played by the Fermi motion, CT effects should
 strongly depend
 upon the value of $x$, so that such a dependence should represent
  a possible signature of CT. Such a  conclusion,
%does not seem
%so convincing upon second thought.
which is certainly well-grounded,  was however  based on the Glauber picture
for the rescattering, in which the hit nucleon
which rescatters  with the other $A-1$ particles
(the {\it ejectile})  is
taken on its mass shell during rescattering. In reality the ejectile
may also
rescatter in a virtual state, below the energy necessary for it
physical excitation.
Thus only a part of the rescattering
contribution vanishes at low energies, namely that coming from the
imaginary part of the ejectile propagator, whereas the  one coming from the
real part does not vanish and may feel the influence of
excited
ejectile states created virtually. Obviously this effect
can be felt
only if one goes beyond the Glauber approximation for
the rescattering.

We have investigated  this problem by a direct calculation
of rescattering contribution in the cumulative region
starting from the relevant Feynman diagrams.
It should be stressed here  that a
 rigorous relativistic description
 in the cumulative region encounters difficulties
of principle already at the level of the impulse approximation; these
difficulties are mainly
related to the treatment of  the  unknown
electromagnetic form-factors for an  off-mass-shell nucleon, so that the
   calculation of the rescattering and FFT contribution with relativistic
spins is out of the question with the present calculational
possibilities.
One has therefore to recur  to some approximations to make
calculations viable. In our approach,  we calculate the amplitudes in a
relativistic manner
treating, however, the deuteron and the nucleons as
spinless particles; the approximation is cured by introducing
  an effective electromagnetic form-factor
of the nucleon suitably chosen to take into account the magnetic moments.
We will demonstrate that at  the level of the IA  such a procedure
provides results in a very good agreement with calculations performed
within a full treatment of spin \cite{CKT}, which makes us confident
that the rescattering and FFT contributions calculated with these effective
form-factors  represents a meaningful approximation.
%This may seem to be too crude an approximation, especially in view
%of numerous calculations done with spinful particles [2].
%The most severe difficulties stem from the treatment
%of spin But in the cumulative region
%a rigorous inclusion of spins only aggravates the matter,
%since it involves the mentioned
%unknown off-mass-shell nucleon electromagnetic form-factors.
%%%%%%%%%%%%%%%%%%%%%%%%%%%%%%%%%%%%%%%%%%%%%%%%%%%%%%%%%%
%To minimize the effects
%of neglecting the spins  and of the the choice of the deuteron wave
%function and nucleon electromagnetic form-factor we study the ratio
%of the FSI to the total  contribution.
%%%%%%%%%%%%%%%%%%%%%%%%%%%%%%%%%%%%%%%%%%%%%%%%%%%%%%%%
Since we are assuming   a quasi-elastic  mechanism  of $\gamma^*$ interaction with a bound
nucleon,  thanks to  the high cumulativity
 $(x>1)$ of the process, the invariant mass
of the final hadronic state does not strongly differ from the
nucleon mass, so that
one  can simply use  elastic rescattering amplitudes for
the rescattering contribution. Most important, in this region the pole
contributions
responsible for the FFT, do not contribute at all and all the FFT effect
originates
from the non-Glauber contribution due to the real part of the ejectile
propagator.
So we shall be able to study directly the importance of this effect
neglected in previous treatments of CT effects.
Our results  show that the relative weight of the
rescattering contribution steadily grows with $x>1$
reaching 50\% already at $x=1.5$. The CT or FFT effects,
indeed
very small at $x<1.3\div 1.4$, become quite pronounced at
larger $x$. So, contrary to usual  expectations, the
study of rescattering on the deuteron allows to see these
effects quite clearly. As already mentioned, our  numerical results
are  obtained in a relativistic approach;  the
approach  based on the Glauber picture
with relativistic kinematics
exhibits no FFT effects, as predicted in  \cite{kope}, but fails
to reproduce the rescattering contribution above $x\sim 1.2$.

Our paper is organized as follows: in Section 2 the general formalism for
treating
inclusive and quasi-exclusive cross sections will be presented; the
Impulse Approximation
will be described in Section 3 whereas the rescattering and FFT effects
will be
treated in Section 4. The numerical results are given in Section 5,
Conclusions are drawn in Section 6.
In Appendices A and B,
several details concerning the numerical
calculations are given.

\section{Kinematics, structure functions and cross-sections}

In the relativistic domain the nucleon struck by the virtual photon
may be excited to states with
higher masses.  At very high energies this leads to the standard picture of
Deep Inelastic Scattering (DIS) in terms
of quarks and gluons. We are interested in this Paper in the quasi-elastic
(q.e.)
scattering, when the final  produced hadronic state is just a nucleon.
To achieve this and stay within the description in terms of nucleons, we
choose the
kinematical region of deep enough cumulativity, $x$ close to 2 ,
where the total mass of the final
hadronic systems remains low, below the threshold for the formation of
excited nucleon states.

  Thus the process we are investigating is as follows

\beq
e+d\to e+p+n
\label{process}
\eeq

\noindent and  in our calculations we simply
use   elastic rescattering amplitudes for
the rescattering inside the deuteron.

We will consider both the inclusive process $d(e,e')np$, when only the
scattered electron is detected in the final state, as well as the exclusive
process $d(e,e'p)n$,
when also the  proton is detected in coincidence with the scattered electron.

The main kinematical quantities  involved in the process will be denoted as follows:
\begin{enumerate}
\item $M$ and $m$ denote  the deuteron and nucleon
 masses, respectively;
\item $k \equiv (E,{\bf k})$ and
$k' \equiv (E', {\bf k}')$ are the
 four-momenta of the  incoming and scattered electrons, respectively;
 \item $q \equiv (q_0,{\bf q})$ is the  photon momentum,
 $Q^2 = -q^2 = 4 E E' sin^2{{\frac{\theta}{2}}}$  the four-momentum
  transfer, $\nu=qp/m$, and  $\theta$ and $\phi$
  the electron scattering
angles in the lab system,  where $\nu =E-E'$;
\item the quantity
\beq
x=\frac{Q^2}{2qp}
\label{bjorken}
\eeq
\noindent is the Bjorken scaling variable  defined with respect  to
the $\gamma^*+N$ scattering, so that
\beq
0\leq x\leq 2
\eeq
\noindent The region $x>1$ is called cumulative;
\item the deuteron momentum is denoted by  $P=2p$;  $k_{1(2)}$ are the momenta of nucleon $1(2)$
before interaction with ${\gamma^*}$ and
 $p_{1(2)}$ are  the
momenta of nucleon $1(2)$ in the final state; 
 
 \item we use light-cone coordinates, $a_+$, $a_-$ defined in the usual way
viz. $a_{+(-)}=(a_0 \pm a_z)/\sqrt{2}$. The independent variables for each
nucleon $i=1,2$ are its transverse momentum component
and the  carried fraction of the momentum $p_+$ ({\it scaling variable}):
$k_{\perp,i}$ and $x_i$ before the interaction and
$p_{\perp,i}$ and $\zeta_{i}$ in the final state. 
Obviously  $x_1+x_2=1$ and $\zeta_1 + \zeta_2 = 2 + \xi$ where
$\xi$ is the  scaling variable  of the photon (see Eq. (\ref{components})
below).

\item in the following we will also need the definition of the nucleon virtuality
 $v$ which is
\beq
v=m^2-k_1^2
\label{virt}
\eeq
for the nucleon before $\gamma^*$ absorption, and
\beq
v'=m^2-(k_1+q)^2
\label{invdopo1}
\eeq
\noindent for the nucleon after absorption.
\end{enumerate}

Since, as it will be shown later on, the exclusive cross section
will be obtained
by a proper procedure from the structure functions appearing in the
definition
of the inclusive cross section, some general relations concerning
kinematics and
structure functions relevant for both inclusive and exclusive processes
will be derived, starting from the definition of the inclusive cross section. In terms
of the standard  structure functions $F_{1(2)}(x,Q^2)$, one has
\beq
%I(E',\theta,\phi)\equiv
\sigma_{incl} \equiv \frac{d\sigma}{dE'd\Omega'}=
\sigma_{Mott}\frac{1}{4\pi M}
\left(\frac{M^2}{qP}F_2(x,Q^2)+2\tan^2\frac{\theta}{2}F_1(x,Q^2)\right)
\label{crossinclu}
\eeq
\noindent
where
\beq
\sigma_{Mott} = \left ( { \alpha \,\,\cos {\theta \over 2} \over
2 { E} \sin^2 {\theta \over 2} } \right )^2
\eeq
is the Mott cross section.
In our normalization the inclusive  structure functions are related to
the imaginary part of the
forward amplitude for the elastic $\gamma^*+d$ scattering (hadronic tensor)
as
\beq
W_{\mu\nu}=\left(-g_{\mu\nu}+\frac{q_{\mu}q_{\nu}}{q^2}\right)F_1(x,Q^2)+
\frac{1}{qP}\left(P_\mu-q_\mu\frac{qP}{q^2}\right)
\left(P_\nu-q_\nu\frac{qP}{q^2}\right)
F_2(x,Q^2)
\label{inclutens}
\eeq
%where the Bjorken scaling variable is
% $q$ and $P=2p$ are the momenta of the photon and deuteron and, as usual
\noindent and the standard way to find the structure functions from
$W_{\mu\nu}$ is to choose
a coordinate system ( the {\it theoretical} system) in which
$q_+=q_y=p_{\perp}=0$.
Labeling  the components of
vectors and tensors in this system  with bars,  one finds
\beq
F_1(x,Q^2)=\bar{W}_{yy},\  \  F_2(x,Q^2)=\frac{Q^2}{4xp_+^2}\bar{W}_{++}
\label{effe12}
\eeq
We take the deuteron at rest, ${\bf p}=0$, but in order
to simplify the calculations of the rescattering contribution, we
choose a system in which  $q_{\perp}=0$ ( the lab frame  for the system
$\gamma^*+d$),
instead of  $q_+=q_y=0$.
These two systems are related by a rotation in the $xz$ plane by the angle
\beq
\phi_0=\arctan\frac{xM}{Q}
\eeq
%where $M$ is the deuteron mass.
 For any vector $k$ we find, in particular
\[
\bar{k}_+=c_+k_++c_-k_-+c_xk_x,\ \ {\bar{k}}_y=k_y,
\]
\beq
c_+=\frac{1}{2}(1+\cos\phi_0),\ \
c_-=\frac{1}{2}(1-\cos\phi_0), \ \
c_x=\frac{1}{\sqrt{2}}\sin\phi_0
\label{kpiumeno}
\eeq
These relations will serve to transform to the lab system
 the structure functions $F_{1,2}$ resulting from our calculation in
 the chosen system.

In the system $q_{\perp}=0$ the longitudinal components of $q$ are
\beq
q_0=\frac{Q^2}{xM},\ \ q_z=-Q\sqrt{\frac{Q^2}{x^2M^2}+1}
\eeq
wherefrom we find the light-cone components
\begin{equation}
q_{\pm}=\frac{Q^2}{xM\sqrt{2}}
\left(1+\sqrt{1\mp \frac{x^2M^2}{Q^2}}\right),\ \
q_+=
%\frac{Q^2}{xM\sqrt{2}}
%\left(1-\sqrt{1+\frac{x^2M^2}{Q^2}}\right)
\xi p_+
\label{components}
\end{equation}
%Here $m$ and $M$ are the  nucleon and deuteron masses respectively.
Evidently $\xi$ is negative and
tends to $-x$ when  $Q^2\to\infty$. Note that in our system the photon
moves along {\it the opposite} direction
of
the $z$-axis.

 Using
the expression (\ref{components})  for the light-cone components of $q$
one obtains for the virtualities given by Eqs. (\ref{virt}, \ref{invdopo1})
\beq
v=2\frac{m^2+k_{1\perp}^2}{x_2}-\frac{1}{2}x_1M^2
\label{invante}
\eeq
and
\beq
v'=Q^2\left(1-\frac{2-x_2}{x}\right)+m_{2\perp}^2\frac{2+\xi}{x_2}
-\frac{1}{2}M^2\left(2-x_2+\frac{1}{2}\xi x_2\right)
\label{invdopo}
\eeq

The total c.m. energy squared for the reaction  $\gamma^*-d$ is
\begin{equation}
s=(2p+q)^2=Q^2\frac{2-x}{x}+M^2
\label{esse}
\end{equation}
As already mentioned,  we shall choose the kinematical domain where formation of
excited
two-nucleon states is impossible or, at least, strongly suppressed, so
that the
reaction $\gamma^*+d$ is essentially binary, i.e. :
\beq
\gamma^*(q)+d(P)\to N(p_1)+N(p_2)
\label{process1}
\eeq
As a result, the  scaling variables and
transverse  momenta of the final nucleons are constrained
by      energy                           conservation
\beq
s=(m^2+p_{1\perp}^2)\left(2+\frac{\zeta_1}{\zeta_2}+
\frac{\zeta_2}{\zeta_1}\right)
\label{encons}
\eeq
The kinematical limits on $\zeta_{1,2}$ are determined
by the condition that  $p_{1\perp}^2>0$ in (\ref{encons}), which leads to
\beq
\frac{1}{2}(2+\xi)\left(1-\sqrt{1-\frac{4m^2}{s}}\right)\leq \zeta_{1,2}
\leq \frac{1}{2}(2+\xi)\left(1+\sqrt{1-\frac{4m^2}{s}}\right)
\eeq
At the limiting values of $\zeta_{1,2}$, we have \quad $p_{1\perp}=0$.

The energy and $z$-component of the momentum of the active nucleon after
absorption of the photon (the {\it fast} nucleon or {\it ejectile}) are given by
\beq
p_{10}=\frac{1}{4}M\left(\zeta_1+
\frac{4(m^2+p_{1\perp}^2)}{\zeta_1M^2}\right),\ \
p_{1z}=\frac{1}{4}M\left(\zeta_1-
\frac{4(m^2+p_{1\perp}^2)}{\zeta_1M^2}\right)
\label{12}
\eeq
They get large if either $p_{1\perp}$ is large or $\zeta_1$ is small.
%In the following, studying the exclusive cross-section for for the
%process given by Eq. \ref{process1},
%we shall limit ourselves to the parallel
%kinematics when $p_{1\perp}=0$. Then the fast nucleon has the
%minimal value of its scaling variable, namely,
%\beq
%\zeta_1=\frac{1}{2}(2+\xi)\left(1-\sqrt{1-\frac{4m^2}{s}}\right)
%\label{zeta1}
%\eeq
%As $s\to \infty$ this value goes to zero. However our calculations will
%be restricted to moderate energies, when the right-hand side of (16)
%is still very different from zero.

To finish this kinematical excursion, note that the scaling variable of
the active nucleon before absorption of the photon  is given by
\beq
x_1=\zeta_1-\xi
\label{xone}
\eeq
which is  different from the scaling variable of the
ejectile nucleon and greater than it (since $\xi$ is negative).
In the high-energy limit $Q^2\to\infty$ $\xi=-x$ and Eq. (\ref{xone})
 transforms into $x_1=\zeta_1+x$. Also
Eq. (\ref{invdopo}) goes into
\beq
v'_{Q^2\to\infty}\simeq Q^2\left(1-\frac{x_1}{x}\right)
\label{invdopoinf}
\eeq
from which the standard relation $x'=x$ follows if the ejectile
lies on the mass-shell.

\section{The Impulse approximation}
\subsection{The inclusive cross section}
In the impulse approximation, depicted in  Fig. 1,  the imaginary part of
the forward $\gamma^*+d$
scattering amplitude is given by
\beq
W_{\mu\nu}=4\gamma^2(Q^2)\int dVK_{\mu}K_{\nu}G^2(v)
\label{incl}
\eeq
where $\gamma$ is the effective electromagnetic form-factor
of the (scalar) nucleon depending only upon $Q^2$, since  we assume
that its dependence on the
virtuality of the proton before the interaction is
effectively taken into account by the deuteron wave function.
The invariant phase volume element is
\beq
dV=
\frac{d^2p_{1\perp}d\zeta_1}{16\pi^2\zeta_1}\delta(m^2-k_2^2)
=\frac{d\zeta_1d\phi}{16\pi^2(2+\xi)}
\label{divu}
\eeq
\noindent where   notations  have been used which will be  convenient for the following
study of rescattering effects, namely  $k_2=p_2=q+2p-p_1$ denotes
the momentum of the
recoil neutron, and, accordingly,  $x_2\equiv \zeta_2$; note, moreover,  that
$k_{2\perp}=-k_{1\perp}=-p_{1\perp}$.

 In Eq.(\ref{incl}), the
 function $G(v)$ is the relativistic deuteron wave function with the
spectator  on the mass shell:
\beq
G(v)=\frac{\Gamma(p,k_1)}{m^2-k_1^2}
\label{wfdeut}
\eeq
where $\Gamma(p,k_1)$ is the vertex function describing
the virtual decay  of the deuteron
into two nucleons, and $v$ denotes the
 virtuality of the active nucleon before interaction,  defined
 previously  (Eq. (\ref{virt}))

In Eq. (\ref{incl}) the four-momentum $K$ has to be chosen to guarantee
conservation of
the electric current from the condition $qK=0$. In our light-cone kinematics
we  find
\beq
K=k_1+yq
\label{Kappa}
\eeq
with
\beq
 y=\frac{qk_1}{Q^2}=\frac{1}{Q^2}\left(Q^2\frac{x_1}{2x}
+\frac{1}{8}x_2\xi M^2-\frac{\xi}{2x_2}m_{2\perp}^2\right)
\label{ipsilon}
\eeq

Two  relevant quantities which we need for the evaluation of the
scattering amplitude
$W_{\mu\nu}$ are:(i)
the virtuality of the off-shell nucleon after interaction defined by
Eq. (\ref{invdopo}),  and (ii)
the components of $W$ in
the theoretical
system; using (\ref{kpiumeno}) we get
\beq
\bar{W}_{++}=4\gamma^2\int dVG^2(v)\Big[c_+K_+
+c_-K_-+c_xK_x\Big]^2
\label{++}
\eeq
\beq
\bar{W}_{yy}=W_{yy}=4\gamma^2\int dV G^2(v)K_y^2
\label{yy}
\eeq

In the lab system neither the longitudinal components of $K$ nor the
virtuality
 $v$ depend
on the azimuthal
angle $\phi$. The $x$ and $y$-components of $K$ are $|p_1|\cos \phi$ and
$|p_1|\sin\phi$ . So the azimuthal
integrations in (\ref{++}) and (\ref{yy})  are trivial: they give an
overall factor
$2\pi$ and an additional 1/2 for the square of transverse components.
We finally obtain
\beq
\sigma_{incl}^{IA}=\frac{\sigma_{Mott}}{4\pi M}\int d\zeta_1
I(|{\bf p}_2|,x,Q^2)
\label {sigia}
\eeq
where
\beq
I(|{\bf p}_2|,x,Q^2,\theta)=\frac{\gamma^2}{2\pi (2+\xi)}
G^2(v)\Big[ 2(c_+K_++c_-K_-)^2+p_{2\perp}^2\left(c_x^2+\tan^2(\theta/2)
\right)\Big]
\label{ia}
\eeq
Here we have taken into account that due to energy conservation,
at fixed $\zeta_1$ (or
$\zeta_2=2+\xi-\zeta_1$) both the longitudinal and transverse components
of the observed and recoiled nucleons become fixed. In fact
$p_{1\perp}^2$ is determined by $\zeta_1$ via Eq. (\ref{encons}). With
$p_{1\perp}^2$ known the longitudinal component
$p_{1z}$ is found using the second of Eqs. (\ref{12}).

Note that $x\sim 1$ and/or small scattering angles $\theta$ the
integrand $I$ is practically independent of $\theta$ and related
only to the internal structure of the deuteron.

In \ref{sigia} one is left only with one non-trivial integration over
$\zeta_1$, which has to be done numerically. This is a big
advantage of the chosen  system: in the theoretical system $v$
results $\phi$-dependent, so that one encounters two numerical
integrations. For the impulse approximation this is still viable,
but passing to rescattering the number of non-trivial integrations
rises to five, which makes the theoretical system impractical.

\subsection{The exclusive cross section}
 We will consider now the  exclusive process (\ref{process}) when  the
the fast nucleon with
momentum $p_1=\{\zeta_1,p_{1\perp}\}$ is observed in coincidence with
the scattered electron.
As is well known, the hadronic tensor  for the exclusive $d(e,e'p)X$ process
has the following general form
\begin{eqnarray}
W_{\mu\nu} &=& \left(-g_{\mu\nu}+\frac{q_{\mu}q_{\nu}}{q^2}\right)
F_1(x,Q^2,p_2)+
\frac{1}{qP}\left(P_\mu-q_\mu\frac{qP}{q^2}\right)
\left(P_\nu-q_\nu\frac{qP}{q^2}\right)
F_2(x,Q^2,p_2) \nonumber \\
&+& {F_{3}(x,Q^2, p_2)} { 1 \over (p_2 \cdot P)}\,
{ 1 \over 2}\,\left (P_\mu p_{2\nu} +
P_\nu p_{2\mu} \right) + {{F_{4}(x,Q^2,p_2)} \over M^2} p_{2\mu} p_{2\nu}
\label{semiincl}
\end{eqnarray}

\noindent i.e. it depends upon four  independent response functions
$F_i$ instead of two appearing in inclusive
scattering. Thus, strictly speaking, it is impossible to obtain the
exclusive cross section from the inclusive
tensor (\ref{inclutens}). However, with all components of $W_{\mu\nu}$
explicitly known in our case, it is easy to demonstrate that by fixing
 the intermediate nucleon scaling variable $\zeta_1$ and
dropping the integration over $d\zeta_1$ in Eq. (\ref{sigia})
one obtains the correct exclusive cross-section integrated over
the azimuthal angles of the final nucleons.  Of course if one
desires to find the exclusive cross-section with a fully fixed
momenta of both the final electron and nucleon, including their
azimuthal directions,  then one has to apply a different procedure
using the explicit expression for the $\gamma^*+d\to p+n$ amplitude
and a convenient kinematical system
instead of representations like (\ref{inclutens}) or (\ref{semiincl}).

\section{The rescattering contribution}
The rescattering amplitude, represented  by the  diagram shown in Fig. 2, has the following form
\begin{eqnarray}
\label{resc}
&&{\cal A}_{\mu\nu} = \\ \nonumber
&&4\gamma^2\int\frac{d^4k_1}{i(2\pi)^4}\frac{d^4{\tilde k}_1}
{i(2\pi)^4}
\frac{K_{1\mu}\tilde{K}_{1\nu}\Gamma(k_1,k_2)\Gamma(\tilde{k}_1,\tilde{k}_2)
a(k'_1,k_2|\tilde{k}'_1,\tilde{k}_2)}
{(m^2-k_1^2)(m^2-k_2^2)(m^2-k^{\prime\,2}_1)(m^2-\tilde{k}_1^2)
(m^2-\tilde k_2^2)(m^2-\tilde k^{\prime\,2}_1)}
 \end{eqnarray}
where $k'_1=k_1+q$,  $\tilde{k}'_1=\tilde{k}_1+q$,  and  $a$ denotes  the
rescattering
amplitude (in the relativistic normalization).
%%%%%%%%%%%%%%%%%%%%%%%%%%%%%%%%%%%%%%%%%%%%%%%%%%%%%%%%%
The first problem one  encounters in the evaluation of Eq. (\ref{resc}),
 is the integration
 over the
"-" components of the momenta (light-cone "energies").
Such an integration requires, in principle,  the knowledge of the unknown
dependence
of both the vertices $\Gamma$ and the rescattering amplitude $a$,
upon the "-" components. In order to overcome such a principle difficulty,
 we make the
approximation consisting of   taking into account only the
singularities coming from the nucleon propagators in Eq. (\ref{resc}).
This  can be justified if both the deuteron wave
function and the rescattering amplitude are  generated by an
instantaneous interaction in the light-cone variables, in which case,
however,
the full relativistic invariance would be lost.
Our procedure is equivalent  to restoring it by
expressing the relativistically invariant arguments via
the light-cone variables in the preferable system where
the interaction was instantaneous.
%%%%%%%%%%%%%%%%%%%%%%%%%%%%%%%%%%%%%%%%%%%%%%%%%%%%%%%%%%%%%%%%%%%%%%%
%Digression on the integration over "-" components in the {\it lab} system
%%%%%%%%%%%%%%%%%%%%%%%%%%%%%%%%%%%%%%%%%%%%%%%%%%%%%%%%%%%%%%%%%%%%%
%%%%%%%%%%%%%%%%%%%%%%%%%%%%%%%%%%%%%%%%%%%%%%%%%%%%%%%%%%

It should be pointed out  that in the lab system with $q_+$ not equal to zero
the integration over the "-" components of the momenta
is more complicated than in the system where $q_+=0$.
Let us choose $k_{2-}$ as an integration variable in the left loop.
The three denominators generate three poles in $k_{2-}$:
\beq
\frac{1}{m_{1\perp}^2/2k_{1+}-P_-+k_{2-}}\quad
\frac{1}{{m'_{1\perp}}^2/2k'_{1+}-P_-q_-+k_{2-}}\quad
\frac{1}{m_{2\perp}^2/2k_{2+}-k_{2-}}
\eeq
where $k'_1=k_1+q$ and all masses are supposed to have a small
negative imaginary part. The position of the poles depends on the signs of
the $"+"$ components. One has to take into account the following
relations, which restrict possible signs of $k_{1,2+}$ and $k'_{1+}$:
\[ k_{1+}+k_{2+}=P_{+}>0,\ \ k'_{1+}=k_{1+}+q_{+}<k_{1+},\ \
k'_{1+}+q_{+}>0\]
following from the fact that the $"+"$ component is positive for physical
particles and negative for the photon.

One sees immediately that if $k_{2+}<0$ or $k_{2+}>P_{+}$ all poles are
on the same side of the real axis in the $k_{2-}$ complex plane, so that
the result of the integration will be zero.
In fact if $k_{2+}<0$ then both $k_{1+}$ and $k'_{1+}$ have to be positive,
and all the poles lie above the real axis.
If $k_{2+}>P_{+}$,  both $k_{1+}$ and $k'_{1+}$ are negative and
all the poles lie below the real axis.

So, as in the system $q_+=0$, we are left with the interval
$0<k_{2+}<P_+$ and consequently also $0<k_{1+}<P_+$.
This means that the pole from the spectator nucleon will lie below the
real axis,  and the one from the active nucleon before the
interaction above it.
However in the lab system there are two different
possibilities for the pole coming from the ejectile propagator:
(i) if $k_{2+}<P_++q_+$ then $k_{1+}>-q_+$ and so $k'_{1+}>0$. This is the
situation which is always realized in the system $q_+=0$, when the
ejectile
pole lies above the real axis and the total integral is given by the
contribution of the residue at the spectator pole at $k_{2}^2=m^2$, (ii)
if $k_{2+}>P_++q_+$ and so $k_{1+}<-q_+$ then $k'_{1+}<0$,
the ejectile pole lies below the real axis, and the integral will then be
given by the residue at the active nucleon pole $k_{1}^2=m^2$.
Thus the result is different for different values of the scaling variables
of the nucleons in the deuteron.
Note that in both regions the integrand, apart from the rescattering
amplitude and
the ejectile propagator, will be expressed via the light-cone function
(\ref{wfdeut})
with one of the nucleons on the mass shell. The
invariant argument $v$ will however be expressed differently in terms
of the integration variables in the described two regions, depending on which
of the nucleons lies on the mass shell.

%%%%%%%%%%%%%%%%%%%%%%%%%%%%%%%%%%%%%%%%%%%%%%%%%%%%%%%%%%%%%%%%%%%
To sum up  we get
\beq
{\cal A}_{\mu\nu}=4\gamma^2\int d\tau d\tilde{\tau}G(v)G(\tilde{v})
K_{\mu}\tilde{K}_{\nu}\frac{a(k'_1,k_2|\tilde{k}'_1,\tilde{k}_2)}
{(m^2-k^{\prime\,2}_1)(m^2-\tilde{k}^{\prime\,2}_1)}
\label{amunu}
\eeq
where
\beq
d\tau=d\tau_2\equiv\frac{d^2k_{2\perp}dx_2}{16\pi^3x_2}\ \  {\rm for}\ \
x_2<2+\xi\ \ (x_1>-\xi)\ \ {\rm Region\ I}
\label{ditau}
 \eeq
and
\beq
d\tau=d\tau_1\ \  {\rm for}\ \  x_1<-\xi\ \ (x_2>2+\xi)\ \
{\rm Region\ II}
\eeq
In Region I, the  virtualities of the active nucleon before and after
interaction and the vector $K$,  are given by
Eqs. (\ref{invante}), (\ref{invdopo}) and (\ref{Kappa}) , whereas
in Region II one has
\beq
v=m^2-k_2^2=2\frac{m_{1\perp}^2}{x_1}-\frac{1}{2}x_2M^2
\label{vu2}
\eeq
\beq
v'=m^2-(k_1+q)^2=Q^2\left(1-\frac{x_1}{x}\right)+
\frac{1}{4}x_1\xi M^2-\frac{\xi}{x_1}m_{1\perp}^2
\label{vprimo}
\eeq
with the four-vector $K$ given by
\beq
 K=k_1+yq
 \label{kappa}
 \eeq
  where
\beq
y=\frac{1}{Q^2}\left(Q^2\frac{x_1}{2x}-\frac{1}{8}x_1\xi
M^2+\frac{\xi}{2x_1}m_{1\perp}^2\right)
\label{yreg2}
\eeq

The c.m. energy squared for the reaction
is given in both regions by Eq. \ref{esse}.
Similar relations hold for variables with tildes in the right integration
loop.

Note that in Region II $ v'=m^2-(k_1+q)^2>0$ and does not vanish.
This means that this region  gives no contribution to the fast nucleon
production.

The  four-momentum transfer $t$ in the rescattering is easily calculated
to be
\beq t=(k_1-\tilde{k}_1)^2=
(x_2-\tilde{x}_2)\left(
\frac{m_{\perp}^2}{x_2}-
\frac{\tilde{m}_{\perp}^2}{\tilde{x}_2}\right)
-(k_2-\tilde{k}_2)_{\perp}^2,\ \ {\rm Region\ I}
\label{ti1}
\eeq
\beq
 t=
(x_2-\tilde{x}_2)\left(\frac{1}{2}M^2-
\frac{m_{\perp}^2}{x_1}-
\frac{\tilde{m}_{\perp}^2}{\tilde{x}_2}\right)
-(k_2-\tilde{k}_2)_{\perp}^2,\ \ {\rm Region\ II}
\label{ti2}
\eeq

In the following we shall introduce a finite formation time for the
rescattering by changing the ejectile propagators in the following way
\beq
\frac{1}{m^2-k^{\prime\,2}_1}\to\frac{1}{m^2-k^{\prime\,2}_1}-
\frac{1}{{m^*}^2-k^{\prime\,2}_1}
\label{denom}
\eeq
where the subtracted term may be considered as an effective
contribution from the
excited ejectile states, which makes the rescattering contribution
vanish at superhigh
energies (equivalent to the colour transparency effect,
see [ 3 ]). In our calculations we have chosen $m^*=1.8$ GeV.

Contributions to $W_{\mu\nu}$ are obtained from (\ref{amunu})
by taking its imaginary part.
The latter contains three terms corresponding to cutting the
initial or final ejectile propagator (terms 1 and 2) and to cutting the
rescattering amplitude. The  contribution to $W_{\mu\nu}$ from the 1st term
comes only from Region I of the integration over $k_2$
(Eq. \ref{ditau}) and is
given by
\[
W_{\mu\nu}^{1,resc}
=4\pi\gamma^2\int d\tau_2d\tilde{\tau}G(v)G(\tilde{v})
\delta(m^2-k^{\prime\,2}_1)
K_{\mu}\tilde{K}_{\nu}\frac{a(k'_1,k_2|\tilde{k}'_1,\tilde{k}_2)}
{m^2-\tilde{k}^{\prime\,2}_1-i0}\]\beq
=4\gamma^2\int dVd\tilde{\tau}G(v)G(\tilde{v})
K_{\mu}\tilde{K}_{\nu}\frac{a(k'_1,k_2|\tilde{k}'_1,\tilde{k}_2)}
{m^2-\tilde{k}^{\prime\,2}_1-i0}
\label{wmn1}
\eeq
where $dV$ is the phase volume (\ref{divu}) of the intermediate real nucleons.

The contribution from the second cut is just the complex conjugate to
Eq. (\ref{wmn1}) with $\mu\leftrightarrow\nu$:
\beq
W_{\mu\nu}^{2,resc}=\Big(W_{\nu\mu}^{1,resc}\Big)^*
\label{cc}
\eeq

We finally come to the third cut, across the rescattering amplitude. It
comes from both Regions I and II and is given by
\beq
W_{\mu\mu}^{3,resc}=4\gamma^2\int d\tau d\tilde{\tau} G(v)G(\tilde{v})
K_{\mu}\tilde{K}_{\nu}\frac{{\rm
Im}\,a(k'_1,k_2|\tilde{k}'_1,\tilde{k}_2)}
{(m^2-k^{\prime\,2}_1+i0)(m^2-\tilde{k}^{\prime\,2}_1-i0)}
\label{wuresc}
\eeq
We shall limit ourselves to the kinematical region of comparatively low
energies where the bulk of the contribution to the imaginary part of the
rescattering amplitude comes from elastic scattering when
\beq
{\rm Im}\,a(k_2|\tilde{k}_2)=\pi\int d\tau_2^{\prime\prime}
\delta(m^2-k^{\prime\prime\,2}_1)
a(k_2|k^{\prime\prime}_2)a^*(k^{\prime\prime}|\tilde{k}_2)
\label{aresc}
\eeq
Putting Eq. (\ref{aresc}) into Eq. (\ref{wuresc}) we get
\beq
W_{\mu\nu}^{3,resc}=4\gamma^2
\int dVd\tau d\tilde{\tau}G(v)G(\tilde{v})
K_{\mu}\tilde{K}_{\nu}\frac{a(k_2|k^{\prime\prime}_2)a^*
(k^{\prime\prime}|\tilde{k}_2)}
{(m^2-k^{\prime\,2}_1+i0)(m^2-\tilde{k}^{\prime\,2}_1-i0)}
\eeq
In this formula one has to take $k^{\prime\prime}=p_1$,
$k^{\prime\prime}_2=q+2p-p_1$ and express
all arguments in terms of light-cone variables of $p_1$.

The total rescattering contribution  can  be represented in a
factorized form via the 4-vector
\beq
X_{\mu}=\int d\tau
K_{\mu}G(v)\frac{a(k^{\prime\prime}_2|k_2)}
{m^2-k^{\prime\,2}_1-i0}
\label{iksmu}
\eeq
where $k^{\prime\prime}_2=2p+q-p_1$ and $k'_1=k_1+q$
>From our formulas we find
\beq
W_{\mu\nu}^{resc}=4\gamma^2\int dV
\Big[G(v)K_{\mu}X_{\nu}+G(v)K_{\nu}X_{\mu}^*+X_{\mu}^*X_{\nu}\Big]
\eeq
Adding to this the contribution from the impulse approximation
(\ref{incl}) we
obtain the total $W_{\mu\nu}$ as an integral of a product of two 4-vectors
\beq
W_{\mu\nu}^{tot}=4\gamma^2\int dV Z^*_\mu Z_\nu
\label{wu}
\eeq
where
\beq
Z_i=G(v)K_i+X_i
\label{zetai}
\eeq
Eq. (\ref{wu}) obviously corresponds to the square modulus of the sum
of two amplitudes shown in Fig.3
%%%%%%%%%%%%%%%%%%%%%%

As in the  the impulse approximation case, the components of
$W_{\mu \nu}$ in the
theoretical system are obtained by means of rotation (\ref{kpiumeno}).
In particular
\beq
\bar{W}_{++}=4\gamma^2\int dV\Big(c_+Z_+
+c_-Z_-+c_xZ_x\Big)^2
\label{dw++}
\eeq
\beq
\bar{W}_{yy}=W_{yy}=4\gamma^2\int dV Z_y^2
\label{dwyy}
\eeq

The longitudinal components of $Z$ do not depend on the azimuthal angle
of the observed nucleon $\phi_1$.
The transverse components have the structure
\beq
Z_\perp=p_{1\perp}\Big(G(v)+\frac{|X_{\perp}|}{|p_{1\perp}|}\Big)
\eeq
where the modulus refers only to vector components ($X$ is complex).
The integrand in the internal azimuthal integration in $X$ only depends
on the
azimuthal angle $\phi_2$ between $p_{1\perp}$ and
 $k_{1\perp}$, which enters the transversal part of momentum transfer
 in the
rescattering.  If the azimuthal angles of $p_{1\perp}$ and $k_{1\perp}$ are
$\phi_1$ and $\phi$, we find
\beq
Z_x=\Big(|K_\perp|G(v)+X_3\Big)\cos\phi_1,\  \
Z_y=\Big(|K_\perp|G(v)+X_3\Big)\sin\phi_1
\label{zxy}
\eeq
where $X_3$ is obtained from $X_\perp$ by substituting
$k_{1\perp}$
by $|k_{1\perp}|\cos\phi$.
Using Eq. (\ref{zxy}) one can trivially do the azimuthal integrations in
(\ref{dw++}) and (\ref{dwyy}).
We obtain in analogy with (\ref{sigia}):
\beq
\sigma_{incl}=\frac{\sigma_{Mott}}{4\pi M}\int d\zeta_1
J(|{\bf p}_2|,x,Q^2,\theta)
\label {sigt}
\eeq
where the integrand is
\beq
J(|{\bf p}_2|,x,Q^2,\theta)=\frac{\gamma^2}{2\pi (2+\xi)}
\Big[ 2|c_+Z_++c_-Z_-|^2+|Z_{\perp}|^2\left(c_x^2+\tan^2(\theta/2)
\right)\Big]
\label{jt}
\eeq
In PWIA $J\to I$ and becomes directly proportional to the deuteron momentum
distribution (c.f. Eq.(\ref{ia})).
The corresponding contribution to the cross-section for the exclusive
process can be obtained by
removing  the integration over $\zeta_1$
%%%%%%%%%%%%%%%%%.

In  Appendix B,  some details concerning the calculation of the vector
$X$,  both in its full relativistic form and within  the approximation of
Glauber-like rescattering, are given. Here
it is instructive to estimate the behaviour of $X$
in the deep cumulative limit. To see this,  we have to put in
our formulas for $v$ and $t$,  $\tilde{x}_2=\zeta_2\to 0$,
keeping $x_2$ generally finite as an integration variable.
Then one finds, neglecting the transverse  motion,
\beq
 t\simeq -\frac{x_2}{\zeta_2}m^2
\label{ti}
\eeq
\noindent with $v$  given by
\beq
v\simeq 2 \frac{m^2}{x_2}
\label{vappr}
\eeq
in Region I and by Eq. (\ref{vu2}) in Region II. Since $\zeta_2\to 0$,
$t\to\infty$ unless $x_2$ also goes to zero. However then
$v\to\infty$ in Region I. Thus,  the behaviour of the
rescattering contribution, apart from the  properties
of the ejectile (including the FFT effect), crucially
depends upon  the relative rate of decrease of
the deuteron wave function and the rescattering amplitude as functions of
their respective invariant arguments.
>From general arguments, such a  behaviour should be similar
in the cumulative limit, since both are generated by the same
interaction. In our calculations however we shall represent
both behaviours via certain phenomenologically chosen functions.
This can be justified up to certain limiting values of
$t$ and $v$ below which these approximations are valid.
However one is not allowed to come
too close to the cumulative limit, where the mutual links
between the asymptotic behaviour of $G(v)$ and $a(t)$
becomes crucial.

\section{Numerical calculations and results}

The central quantity we have to calculate is the vector $X_{\mu}$ given
by (\ref{iksmu}).
According  to Eq. (\ref{ti2}), when FFT is taken into account, this vector
 contains two terms
 differing  in the propagator of the ejectile. Both terms are
calculated in a similar manner, the only difference being  the mass of
the ejectile. Some details of the numerical evaluation of   $X_{\mu}$
are given
in Appendix B. The basic ingredients entering the definition of
$X_{\mu}$ are the
relativistic   deuteron wave function $G(v)$, the elastic scattering
amplitude $a$,   and the effective
form factor $\gamma$ appearing in the expression of the hadronic tensor.
For the relativistic deuteron wave function
$G(v)$,
we have taken a relativistic generalization of the
non-relativistic function
$\Psi({\bf k}_1^2)$ 
which corresponds to the $AV14$ interaction \cite{interaction}.
Its relation to $G(v)$ was established from the
non-relativistic limit (see Appendix A), obtaining:
\beq
G^2(v)=2M(2\pi)^3)|\Psi({\bf k}_1^2)|^2
\label{Gi}
\eeq
\noindent with
\beq
{\bf k}_1^2=\frac{1}{2}(v-M\epsilon)
\label{kap}
\eeq
As for the rescattering amplitude, it was chosen  in the form
\beq
a(s,t)=(\alpha+i)\sigma^{tot}(s)\sqrt{s(s-4m^2)}e^{bt}
\eeq
with the values of the parameters  $\sigma^{tot}$, $\alpha$ and $b$
taken from \cite{amplitude}.

The effective nuclear form-factor $\gamma(Q^2)$ was chosen to take into
account the magnetic interaction of realistic nucleons. A comparison of
our impulse approximation results
with the results obtained taking spin into accounts
in the region of small cumulativity $x\sim 1$ leads to the choice
\beq
\gamma^2(Q^2)=\gamma_D^2(Q^2)\frac{2+\tau (\mu_p^2+\mu_n^2)}{1+\tau}
\label{formfactor}
\eeq
where
\beq
\gamma_D(Q^2)=\left(1+\frac{Q^2}{0.71 (GeV/c)^2}\right)^{-2}
\eeq
is the dipole ($D$)form-factor,
$\tau=Q^2/(4m^2)$, and $\mu_{p,n}$ are the anomalous magnetic moments of
the proton and neutron.

To quantify the effects of the non-Glauber nature of the relativistic
rescattering
we also repeated the calculations taking for the rescattering the Glauber
form.
This involves  two approximations:
\begin{enumerate}
 \item first, the longitudinal part
of the
momentum transfer in the rescattering is disregarded, which means that
\beq
t=-(k_2-\tilde{k}_2)_{\perp}^2
\eeq
\item  second, and most important in our case,  the contribution
from the real part of the virtual ejectile propagator is neglected, and only
the full contribution from
its pole is taken into account.
\end{enumerate}

 Within
 the above  approximations, evidently,  FFT effects vanish unless the
rescattering
energy becomes greater than the threshold for  the production of the
excited ejectile
with  mass $m^*$. Our energies lie below this threshold, so that within
 Glauber rescattering we find  no FFT effects.

 \subsection{The inclusive d(e,e')np cross section}

%\section{Numerical results}
The cross-section for the inclusive process $d(e,e')pn$ is given by
Eq. (\ref{sigt}), with
$J$ given by Eq. (\ref{jt}).
Our numerical calculations have been performed in correspondence of the
experimental  data of \cite{expincl} with initial electron energy
$E=9.761$ GeV and  scattering  angle $\theta=10^o$. We have considered
 values for the final electron energy which cover the region of $x$
in the interval $1.0<x<1.71$. Some relevant kinematical values characterizing   the
chosen points are listed in Table 1. It can be seen, as already pointed out
in Ref. \cite{CKT}, that at high
values of $x$ (high cumulativity) the value of $p_{lab}$, representing the momentum of the
hit nucleon in the {\it lab} system of a second nucleon,  is very small, well below
the threshold energy
for pion production.
This justifies our approach  in terms of nucleon degrees of freedom only,
which means that
only  the elastic
rescattering amplitude is used  and the
excited nucleon states are omitted.

 The IA results calculated within our
approach containing  the effective form factor given by Eq.
(\ref{formfactor}) is compared in Fig. 4
 with the  full calculation
of Ref. \cite{CKT} where 
particle spins  are   correctly taken into account; it can be seen that
the two approaches
yield practically the same results. Thus we are confident that the use
of spinless particles with effective
form factors to take into account  rescattering and FFT effects,
is a significant one.

The effects of rescattering and FFT on the inclusive cross section
are shown in Figs. 5 and 6 (from now-on, unless differently stated,
$FSI$ is used to mean
that {\it both}  rescattering and FFT effects are taken into account).
In Fig. 5 we show the ratio of the cross section which includes $FSI$
to  the $PWIA$ cross section.
 It can be seen  that
that rescattering contributions, which are  very small at
$x\sim 1$ (corresponding to $\nu \simeq 1.332\,\, GeV$, cf. Table 1),
 steadily grows with  $x$, reaching an
order of about 50\% already at $x\sim 1.3$.
In Fig. 6 and 7 the theoretical results are compared with the
experimental data.
It can be observed that, up to $x\sim 1.4$ the
 the cross-section which includes rescattering effects
 is close to the experimental one at $x\sim 1$
but becomes somewhat lower towards $x=1.4$. Up to $x=1.2$ the FFT effects
are negligible but starting from this value tend to substantially diminish
the cross-section, whereas above $x=1.4$ the situation abruptly changes:
the rescattering  strongly enhances the cross-section, an effect which is
common to all  approaches which take into account FSI (see e.g. Ref.
\cite{FSIincl}); the effect was found particularly relevant in finite
nuclei and various phenomenological effects have been adopted to contrast
rescattering effects; from the results presented in Fig. 5, 6 and 7, it
can be seen
that FFT effects, decreasing the cross section for an off-shell nucleon,
lead precisely to the desired effect: when FFT are considered,
the cross section is again reduced toward the experimental data.
 The Glauber approximation
seems to work quite well at low cumulativity, $x<1.1$ ($\Delta E>1.2$
GeV). At higher $x$ the Glauber rescattering results are very different
from the ones with a full relativistic treatment, both in sign and
magnitude. In fact the rescattering in the Glauber approximation gives a
relatively
 small contribution, so that the total cross-section becomes
quite close to the impulse approximation.
It can be seen that rescattering and FFT effects are necessary
 in order to improve  the agreement at high cumulativity but they tend
 to destroy
 the better agreement between experimental data and the PWIA.

%%%%%%%%%%%%%%%%%%%%%%%%%%%%%%%%%%%%%%%%%%%%%%%%%%%%%%%%%%%
%As for  the absolute values of the cross-sections obtained
%under various approximations one once again sees
%the rescattering and FFT effects  clearly observable at $x>1.4\div 1.5$.
%We have also shown the cross-sections calculated with the Glauber
%approximation for the rescattering. As one observes,
%this approximation seems to work quite
%well at low cumulativity, $x<1.1$. At higher $x$ the Glauber
%rescattering contribution
%results rather small, so that the total cross-section becomes
%quite close to the impulse approximation.
%%%%%%%%%%%%%%%%%%%%%%%%%%%%%%%%%%%%%%%%%%%%%%%%%%%%%%%%%%%%
\subsection{The exclusive d(e,e'p)n cross section}
 According to (\ref{sigt}) the cross section for  the exclusive process
$d(e,e'p)n$ has the following form
\beq
\sigma _{excl}\equiv\frac{d\sigma}{d\Omega' dE' d\zeta_1}=
\frac{\sigma^{Mott}}{4\pi M} J(|{\bf p}_2|, x, Q^2,\theta)
\label{exclusive}
\eeq
where the definition of $ J$
is given by  Eq. (\ref{jt}).
We will consider the reduced cross section i.e..
\beq
n_{eff}=|\Psi(|{\bf p}_2|)|^2\frac{\sigma_{excl}}{\sigma_{excl}^{PWIA}}=
|\Psi(|{\bf p}_2|)|^2\frac{J(|{\bf p}_2|, x, Q^2,\theta)}
{I(|{\bf p}_2|, x, Q^2),\theta}
\label{ratio1}
\eeq

\noindent which, in  PWIA turns into  the nucleon momentum distribution.
\beq
n(|{\bf p}_2|)=|\Psi(|{\bf p}_2|)|^2
\label{neff}
\eeq
 
As already  mentioned, the  kinematics chosen in the present paper corresponds
to small scattering angles  $\theta$
l, so that  $\theta$-dependence of both $J$ and $I$
coming from  terms proportional to $\tan^2(\theta/2)$ is negligible.

 Figs. 7-11 show the effective momentum distributions calculated taking
 rescattering and FFT  into account.
  Calculations have been performed  
   fixing the
 values of  $x$ and $Q^2$ and  varying the missing momentum
 $|{\bf p}_{mis}| = |{\bf q} - {\bf p}_1|=
|{\bf p}_2|$ which means that, according to the energy conservation of
the process,
the angle  between  $|{\bf q}|$ and  $|{\bf p}_2|$
changes for every value of  $|{\bf p}_2|$.\
 We have first considered the process at the top of the quasi-elastic peak
($x=1$) and then  in the cumulative region ($x>1$).
For each value of $x$, calculations have been performed
in correspondence of three values of $Q^2$, viz. $Q^2 =  2, \,\,5$
and $10
 (GeV/c)^2$.

The results corresponding to $x=1$ are shown in Figs. 7 and 8, whereas
the ones corresponding to
$x=1.8$ are presented in Figs.\,\,9 and 10.
In Fig.7  the nucleon momentum distribution $n(|{\bf p}_2|) \equiv
n_{eff}^{PWIA}(|{\bf p}_2|)$
is  compared with $n_{eff}(|{\bf p}_2|) \equiv n_{eff}(|{\bf p}_2|,x,Q^2)$
obtained at various values of $Q^2$, including both rescattering an FFT
effects. The results at  $Q^2=0.5 (GeV/c)^2$  show, in agreement with
other calculations (see e.g.
\cite{artmut}), that FSI effects which are very small
at $p_2 \simeq 0$, lead to an appreciable increase
of $n_{eff}$ at $p_2 \geq  0.2 GeV/c$ 
as also recently confirmed
by new experimental data  ( \cite{ulmer}).
 Fig. 8 illustrates 
 FFT effects, which are given
by the difference between the full curves, which include {\it both}
rescattering and FFT effects, and the dashed curves,  which include only
rescattering effects. It can be seen that the FFT   increases
with $Q^2$ but,
 at large values of $Q^2$, it seems  to decreases,
in agreement with the results obtained for  the process $^4He(e,e'p)^3H$
(\cite{FFTHiko}). The results at $x=1.8$ are presented in Figs. 9 and 10.
Because of kinematical restrictions imposed by energy conservation,
only limited
ranges of variation of $p_2$ are allowed. Fig. 9 shows
that with increasing momentum transfer FSI effects decrease in
an appreciable way which is due, as illustrated in Fig. 10, to
FFT effects, which makes
the distorted cross-section more similar to the PWIA one. Note that in the
cumulative region, the effective momentum distribution
are  lower than the  the PWIA results,
whereas at $x=1$ one observes the opposite effect.
In Fig. 11  $n_{eff}$ at $x=1$ and $x=1.8$ are shown together to
illustrate  how
FSI decrease with $Q^2$. Eventually,
in Fig. 12, our rescattering results are compared with the Glauber results.
It can be seen that
with increasing momentum transfer the Glauber approximation provides
very poor results, which
 is particularly true
in the cumulativity region. In order to better understand the  the difference between the
effects of rescattering and and FFT effects, , in Fig. 13 we show  the
 Transparency $T$, defined as
\beq
T=\frac{n_{eff}^{FSI}(|{\bf p}_2|, x, Q^2)}{n_{eff}(|{\bf p}_2|)}
\label{transp}
\eeq
where $n_{eff}^{FSI}$ includes both the rescattering and FFT terms
(full lines) or only
the rescattering contribution (dashed lines). It can be seen, as
already pointed out, that
 FFT effects lead to opposite results  at $x=1$ and in the
 cumulative regions, respectively,
 at $x=1$ FFT effects decrease the overall contribution of FSI.
In Ref. \cite{mark}, the process $D(e,e'p)n$ has been calculated by
treating final state
 rescattering
within the Glauber approach and taking into account
colour transparency effects by a quantum diffusion model
and by a three-channel approach. Although a direct comparison with the
results of Ref. \cite{mark}
is in principle difficult due to the fact that there the variable
$\zeta_2$ ($\alpha$,
in the notation of
Ref. \cite{mark}) is fixed at  $\zeta_2 = 1$, whereas in our kinematical
conditions all values of $\zeta_2$  allowed by energy conservation
are included, a qualitative agreement between
the two calculations can be observed.

\section{Conclusions}

By using a relativistic approach based upon the evaluation  of
Feynman diagrams, we have calculated
the rescattering contribution to the  cross-sections of the inclusive,
$d(e,e')np$, end exclusive,   $d(e,e'p)n$, processes, paying
particular attention to the
so-called cumulative region, i.e. the region corresponding to $x>1$.
Besides the $p-n$  rescattering in the final state,
we have also considered  colour transparency
effects by introducing the finite formation time of the hit hadron;
such an approach, which is simple and
 physically well grounded,
 differs from the ones used  previously
mainly in that the virtuality of the hit nucleon is explicitly taken
into account.
To make calculation viable, two main approximations have been adopted,
namely
\begin{enumerate}
\item the nuclear vertex function has been obtained by relativization of the
non relativistic
deuteron wave function corresponding to the $AV14$ interaction;
\item spin-less nucleons  have been considered, but an effective nucleon form factor has been introduced
to take into account magnetic moments effects.
\end{enumerate}
Both approximations have been carefully investigated: as far as the first one is
 concerned, relativistic effects in the nuclear wave function
appear to be relevant only at high values of $p_2$ and $Q^2$ and play
a minor role at $p_2 \leq 0.6 (GeV/c)$ and $Q^2 \leq
5 (GeV/c)^2$, which is the main region of our investigation;
as for the second one,
we obtained a good agreement, within the PWIA,
between our approach and the  results obtained taking spin nucleon
spin correctly into account, which makes us confident
in our treatment of rescattering and FFT effects.

Concerning the main results we have obtained, they can be
summarized as follows:
\begin{enumerate}
\item As shown in Figs. 5 and 6, rescattering effects in the
inclusive cross section  appreciably increase
 with $x$.   At $x\sim 1.4 \div
1.5$ they decrease the cross section with respect to
the PWIA,
in some disagreement with the experimental data. At such values of
 $x$ the FFT effects  reduce  the magnitude
of rescattering, which changes its relative weight
in a very pronounced manner due to  interference with the
impulse approximation contribution.  At still larger values of $x$
rescattering increases the cross-section bringing it into better
agreement with the data.
This phenomenon may serve to search for FFT or CT
effects in the rescattering at comparatively small $Q^2$.
 
\item our results show that at $x=1$ rescattering effects on the
exclusive cross section
 $d(e,e'p)n$ become very important at $p_2 \geq 0.2 GeV/c$ and
 exhibit a strong $Q^2$ dependence (cf. Fig. 1), which may serve
 a significant signal about the nature of rescattering effects.
 In the cumulative region rescattering effects, at the same value of
 $Q^2$ appear to be less (cf.
 Fig. 10). As for FFT effects, illustrated in Figs. 8 and 10,
 they start to be important only at
 $Q^2 > 5 GeV/c$  and high values of $p_2$; in particular,
 at $x>1$ and $Q^2 = 10 GeV/c$, they give at $p_2 \simeq 1.6
 GeV/c$ a contribution of about an order of magnitude (cf. Fig. 10).
 \item  we have shown that  although the Glauber approximation works
 quite well at low
cumulativity, at high values
of  $x>>1$ it fails to reproduce our  relativistic
rescattering both quantitatively and qualitatively.
\end{enumerate}

In conclusions, the results exhibited in this paper demonstrate that 
the diagrammatic approach we have developed can be successfully applied to the two body case; 
the application to complex nuclei does not require additional difficulties: it is underway and the results will be presented 
elsewhere \cite{complex}. 

\section{Acknowledgments}
We are grateful to B. Kopeliovich for reading the manuscript and
useful suggestions. M.A.B and L.P.K. are deeply thankful to the INFN and
University of Perugia (Italy) for hospitality and financial support.
This work was partly supported by the RFFI Grants 01-02-17137 and 
0015-96-737 (Russia).

\appendix
\section{Some details on the numerical calculations}
The central quantity we have to calculate is $X_{\mu}$ given
by (\ref{iksmu}).
According  to (\ref{denom}), when FFT is taken into account,  it is
a difference of two terms
\beq
X_{\mu}=X_{\mu}^{(1)}-X_{\mu}^{(2)}
\eeq
which differ in the propagator of the ejectile. Both terms are
calculated in a similar manner, the only difference being  the mass of
the ejectile.
 In the following we discuss the first term, not
specifying it explicitly.

The components of the vector $K$ (Eq. (\ref{kappa})) have the following
values
\beq
K_+=p_+\left(2-x_2+y\xi\right),\ \
K_{\perp}=-k_{2\perp}
\eeq
and
\[
K_-=2p_--\frac{m_{2\perp}^2}{2x_2p_+}+yq_-,\ \ {\rm Region\ I}
\]\beq
K_-=\frac{m_{1\perp}^2}{2x_1p_+}+yq_-,\ \ {\rm Region\ II}
\eeq
where $y$ is given by Eqs. (\ref{ipsilon}) and (\ref{yreg2}) in
Regions I and II respectively.

At a given  transverse momentum of the real fast nucleon $p_1$,
we direct it along the
$x$-axis. Then the component $K_{3}$ in $X_{3}$ introduced in the
previous section  will evidently be
\beq
K_3=|k_{2\perp}|\cos\phi\equiv k\cos\phi
\eeq
where we denote $k=|k_{2\perp}|$
The rescattering amplitude integrated over
the azimuthal angle will give a function
\beq
b_\mu(x_2,k^2)=\int_{0}^{2\pi}d\phi  a(s,t) \cos^{m}\phi
\eeq
where $m=0$ for $\mu=\pm$, $m=1$ for $\mu=3$
 and $t$ is determined
according to Eqs. (\ref{ti1}) and (\ref{ti2}) in which we have to put $\tilde{x}_1=\zeta$ and
$\tilde{k}_{1\perp}=p_{1\perp}$.

So after the integration over the azimuthal angles we get
\beq
X_{\mu}=\frac{1}{32\pi^3}\int_0^2\frac{dx_2}{x_2}\int_0^{\infty}
dk^2 K_{\mu}G(v)b_\mu(x_2,k^2)\frac{1}{m^2-(k_1+q)^2-i0}
\eeq
where $\mu=\pm,3$, $K_{3}\to k$,
 $v$ is given by Eqs. (\ref{invante}) and (\ref{vu2}) and the denominator is
 given by Eqs. (\ref{invdopo})
and (\ref{vprimo}).

We put
\[ k^2=\frac{w}{1-w},\ \  w=\frac{k^2}{1+k^2},\ \  dk^2=\frac{dw}{(1-w)^2}\]
to convert the integration region into the interval [0,1].
To separate the singularity in Region I we present the denominator
(\ref{invdopo}) in the form
\beq
m^2-(k_1+q)^2-i0=\frac{2+\xi}{x_2}(k^2-\lambda-i0)
\eeq
where
\beq
\lambda=-\frac{x_2}{2+\xi}\Big[Q^2\frac{x+x_2-2}{x}+m^2\frac{2+\xi}{x_2}-
\frac{1}{4}M^2(4-2x_2+\xi x_2)\Big]
\eeq
The singularity is present only if $\lambda>1$. For such $\lambda$
we transform the integral  over $k^2$  into
\beq
\frac{x_2}{2+\xi}\int_0^1\frac{dwf(w)}{w-w_0-i0}\equiv\frac{x_2}{2+\xi}L
\eeq
where
\beq
f(w)=\frac{\bar{K}_\mu G(v)b_{\mu}(x_2,k^2)}{(1-w)(1+\lambda)}
\eeq
considered as a function of $w$ at fixed $x_2$
and
\beq
w_0=\frac{\lambda}{1+\lambda}
\eeq
If
 $\lambda>0$ then so $0<w_0<1$. We  present the integral $L$ as
\beq
L=\int_0^1\frac{dw(f(w-f(w_0))}{w-w_0-i0}+f(w_0)
\left(\ln\frac{1-w_0}{w_0}+i\pi\right)
\eeq
The left integral over $w$ has no singularity at $w=w_0$ and can be
calculated numerically. After that the left integration over $x_2$
presents no difficulties in principle.

\section{ Relation  of $G(v)$ to the non-relativistic
wave function of the deuteron}

We shall find this relation studying the non-relativistic limit.
Comparing the diagrams with a direct interaction of the soft photon with the
deuteron and via its proton constituent we find
\beq
2P_\mu=\int \frac{d^4k_1}{(2\pi)^4i}2k_{1\mu}G^2(v)\frac{1}{m^2-k_2^2}
\eeq
Taking the zero component and integrating over $k_{20}$ we find
\beq
2M=\int \frac{d^3k_2}{(2\pi)^3}2k_{10}G^2(v)\frac{1}{2k_{20}}
\eeq
In the nonrelativistic limit $k_{10}=k_{20}=m$ so that we get
\beq
\int \frac{d^3k_2}{2M(2\pi)^3}G^2(v)=1
\eeq
wherefrom we conclude that
\beq
G^2(v)=2M(2\pi)^3\Phi({\bf k}_2^2)
\eeq
(of course ${\bf k}_2^2$ can be substituted by ${\bf k}_1^2$ since
${\bf k}_1+{\bf k}_2=0$)

\begin{table}[h,t]

\begin{center}
\caption{}
\vskip 2mm

\begin{tabular}{|c|c|c|c|c|}\hline
           &        &      & & \\
$\nu$, GeV &    x   & $Q^2,\, (GeV/c)^2$ &$s,\, GeV^2$ &$p_{lab},\,GeV/c$ \\
           &        &      &    &     \\\hline
0.826      &1.71    &2.65  &3.96   & 0.73    \\
0.872      &1.61    &2.64  &4.14   & 0.88    \\
0.930      &1.50    &2.62  &4.38   & 1.06    \\
0.987      &1.41    &2.60  &4.61   & 1.21    \\
1.056      &1.30    &2.58  &4.89   & 1.40    \\
1.137      &1.20    &2.56  &5.22   & 1.61   \\
1.228      &1.10    &2.53  &5.58   & 1.82   \\
1.332      &1.00    &2.50  &6.00   & 2.07   \\\hline
\end{tabular}
\end{center}
\vskip 2mm
{Some kinematical variables relevant to the experimental data considered in the present paper:
 $\nu$ is the energy transfer, $x$ the Bjorken scaling variable,
$Q^2$ the square of the 4- momentum transfer, $s=2m^2+2m \sqrt{p_{lab}^2+m^2}$
the  two-nucleon invariant mass  in the final state, and  $p_{lab}$ the momentum of the struck nucleon in
the final state in the {\it lab} system of the spectator nucleon. Note that    the inelastic
channel contributions start to be relevant at $p_{lab} \simeq 1.2\, GeV/c$.}
\label{table1}
\end{table}
%%%%%%%%%%%%%%%%%%%%%%%%%%%%%%%%%%%%%%%%%%%%%%%%%%%%%%%%%%%%%

\newpage
\begin{figure}[ht]
\epsfxsize 4in
\centerline{\epsfbox{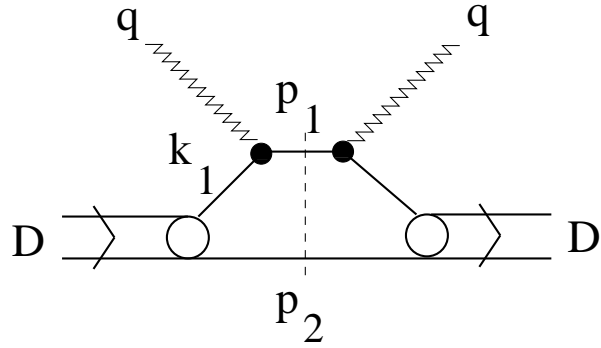}}
\caption{The forward  $\gamma^*$- d scattering amplitude corresponding to the
 Plane Wave Impulse Approximation.}
\label{Fig1}
\end{figure}
\begin{figure}[ht]
\centerline{\epsfbox{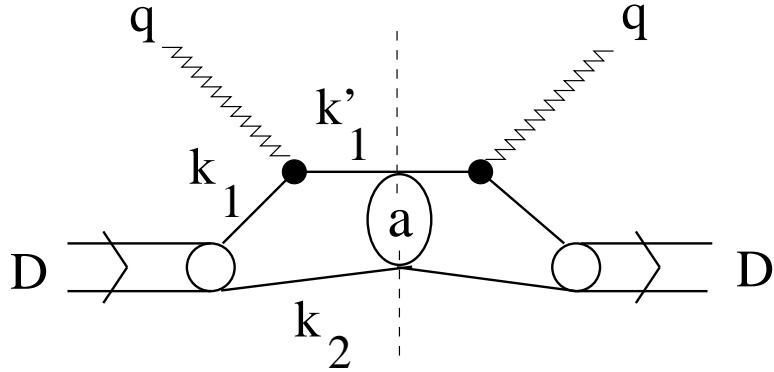}}
\caption{ The forward  $\gamma^*$- d scattering amplitude corresponding to the $p-n$ rescattering in
the final state.}
\label{Fig2}
\end{figure}
 \begin{figure}[ht]
\centerline{\epsfbox{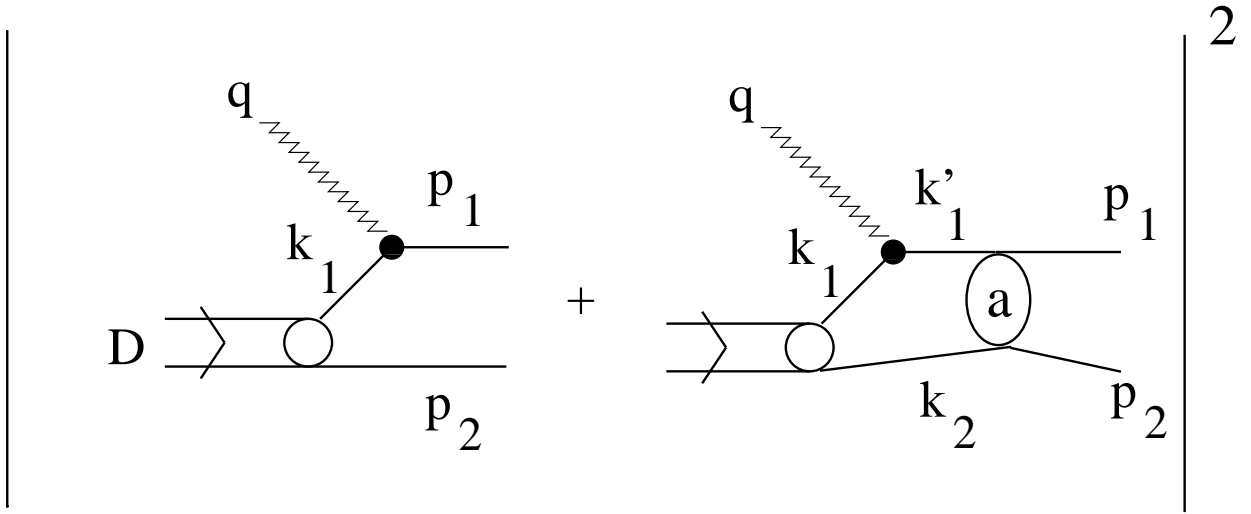}}
\caption{ The square of the sum of
the impulse approximation and the  rescattering amplitudes which govern the inclusive and exclusive cross sections.}
\label{Fig3}
\end{figure}
\begin{figure}[ht]
\epsfxsize 3.5in
\centerline{\epsfbox{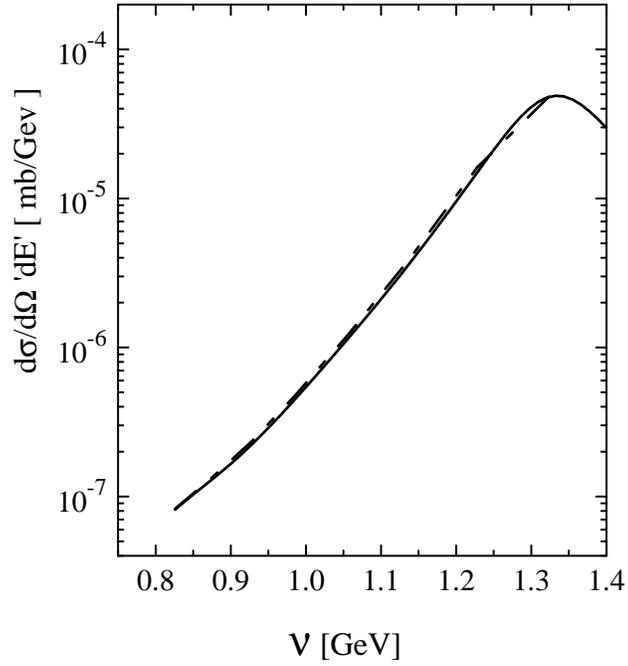}}
\caption{The inclusive cross section
calculated in PWIA. {\it Full curve}: results of Ref. \cite{CKT}  obtained taking nucleon spin correctly into account;
{\it dashed curve}: present results based upon spin-less nucleons with the effective form factor given by Eq. \ref{formfactor}.
 In this
Figure, as in Figs. 5 and 6, the value of $x$ corresponding to a given value of $\nu$ is listed in Table 1}
\label{Fig4}
\end{figure}
\begin{figure}[ht]
\epsfxsize 3.5in
\centerline{\epsfbox{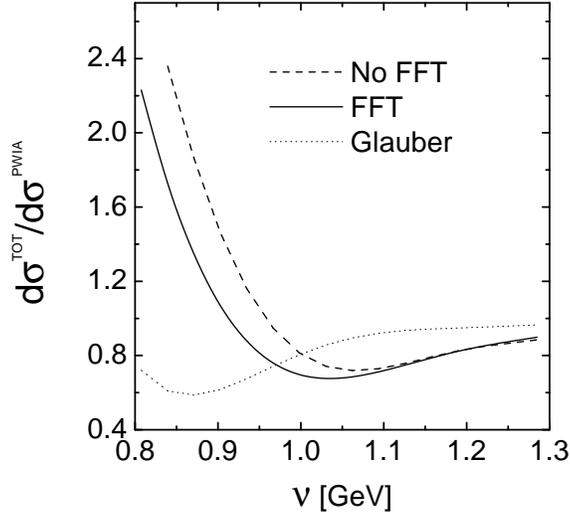}}
\caption{ Ratio of the  cross section $d\sigma^{FSI} \equiv \frac{d\sigma}{dE'd\Omega'}$ to
 the PWIA cross section $d\sigma^{PWIA} \equiv\frac{d\sigma}{dE'd\Omega'}$.
 The dashed curve corresponds do $d\sigma^{tot}$ which includes only rescattering
 effects, whereas the full curve includes both rescattering and FFT effects.
 The dotted curve corresponds to the Glauber approximation for the rescattering.}
\label{Fig5}
\end{figure}
\begin{figure}[ht]
\epsfxsize 3.5in
\centerline{\epsfbox{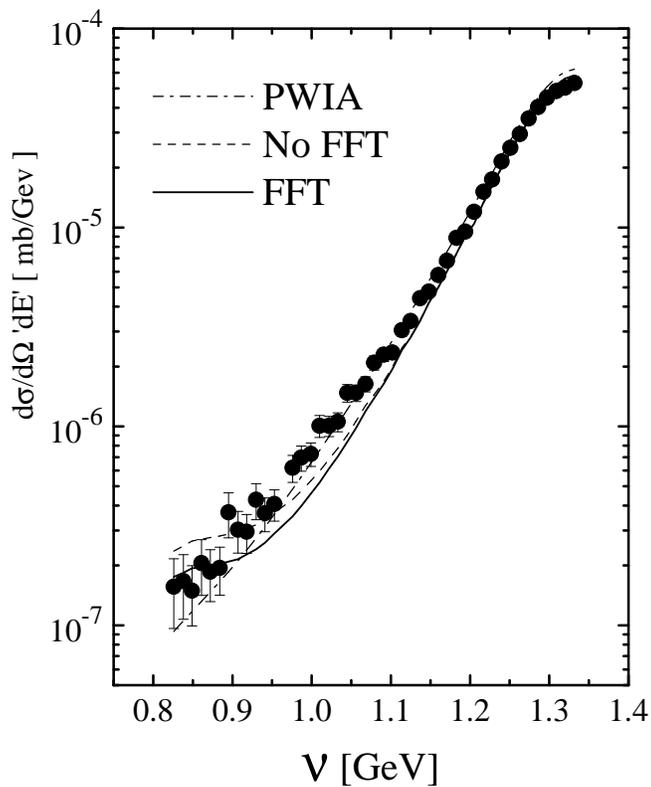}}
\caption{Comparison of the  experimental data from Ref. \cite{expincl} ($E=9.761 GeV$) with the results of our calculations
corresponding to the PWIA (dot-dash), PWIA plus rescattering (short-dash) and PWIA plus rescattering  and  FFT (full).}
\label{Fig6}
\end{figure}
\begin{figure}[ht]
\epsfxsize 3.5in
\centerline{\epsfbox{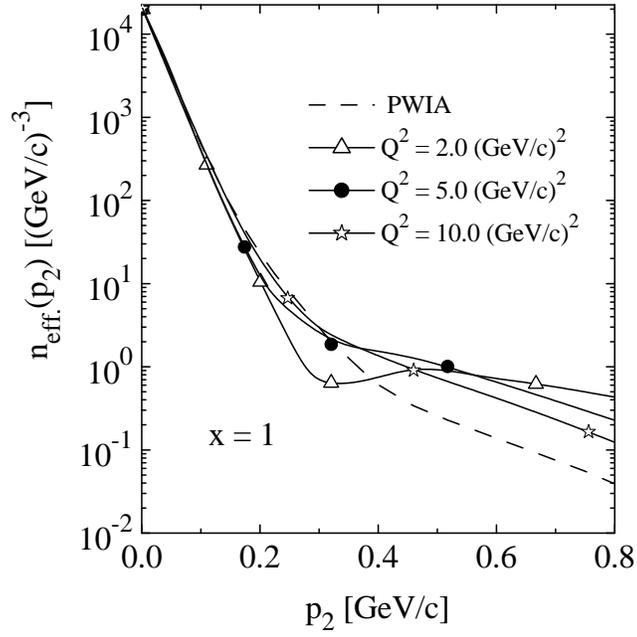}}
\caption{ The effective momentum distribution
(Eq. \ref{ratio1})  {\it vs.}
 the neutron recoil momentum ${|\bf p}_2| \equiv p_2$ at $x=1$ . The dot-dashed curve represents the PWIA result,
 whereas the other curves include also rescattering and FFT effects at various values of  $Q^2$. In this Figure,
 as in Figures
 8-11,  $n_{eff}(|{\bf p}_2|) \equiv n_{eff}(|{\bf p}_2|,x,Q^2)$.}
\label{Fig7}
\end{figure}
\begin{figure}[ht]
\epsfxsize 3.5in
\centerline{\epsfbox{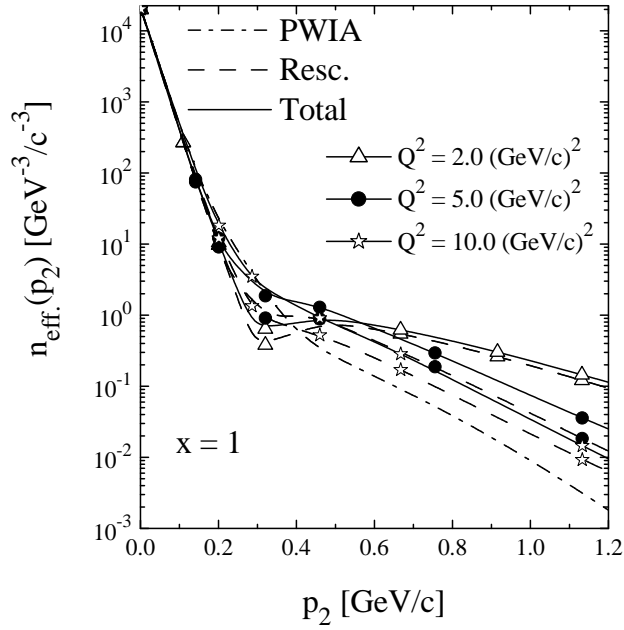}}
\caption{ The effective momentum distribution
(Eq. \ref{ratio1})  at $x=1$  which includes
rescattering and FFT effects (full),  and rescattering effects only (dashed); the difference
between  the full and dashed curves is due to FFT effects; the dot-dashed curve represents the
PWIA result.}
\label{Fig8}
\end{figure}
\begin{figure}[ht]
\epsfxsize 3.5in
\centerline{\epsfbox{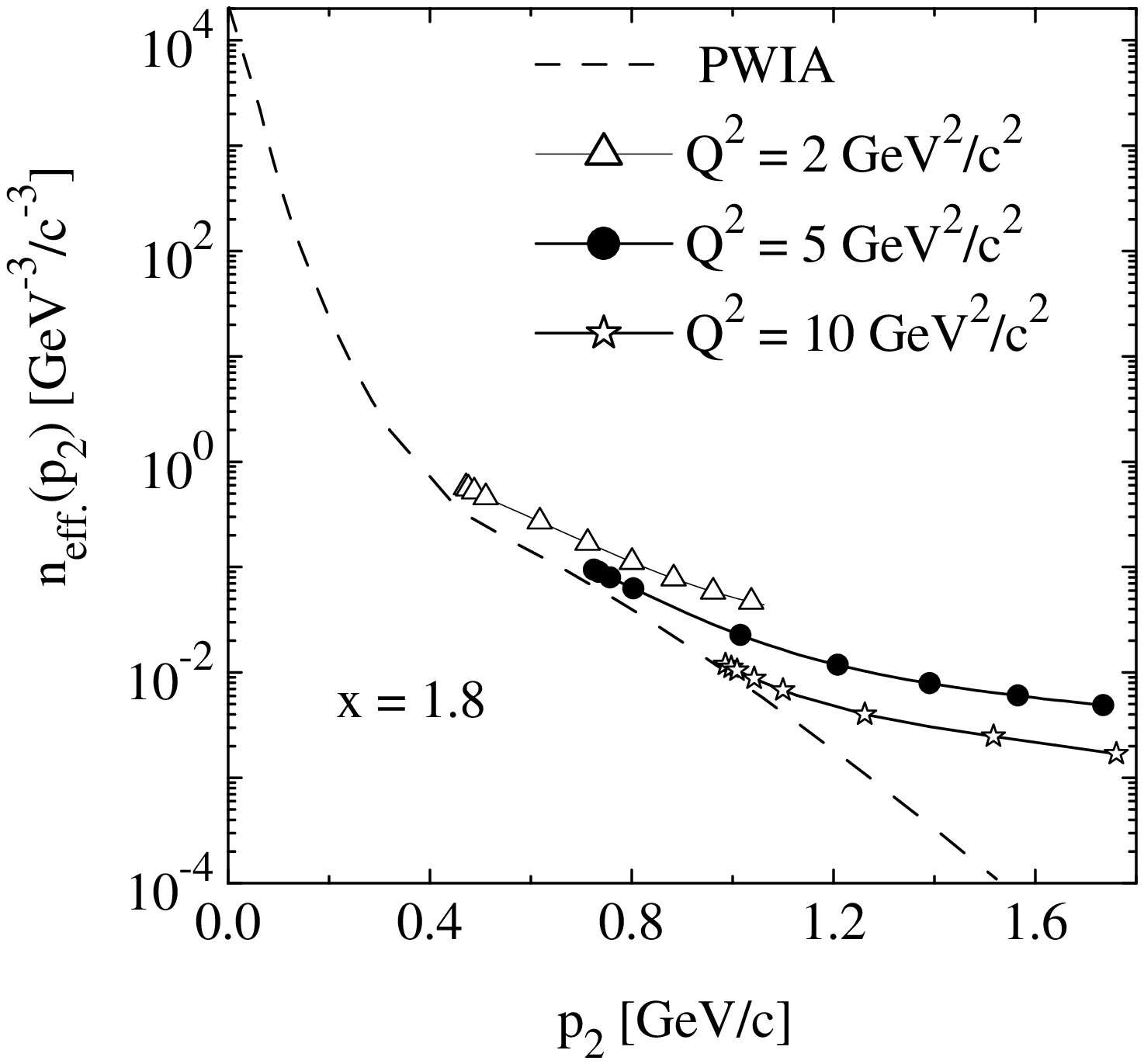}}
\caption{ The same as in Fig. \ref{Fig7} at $x=1.8$.}
\label{Fig9}
\end{figure}

\begin{figure}[ht]
\epsfxsize 3.5in
\centerline{\epsfbox{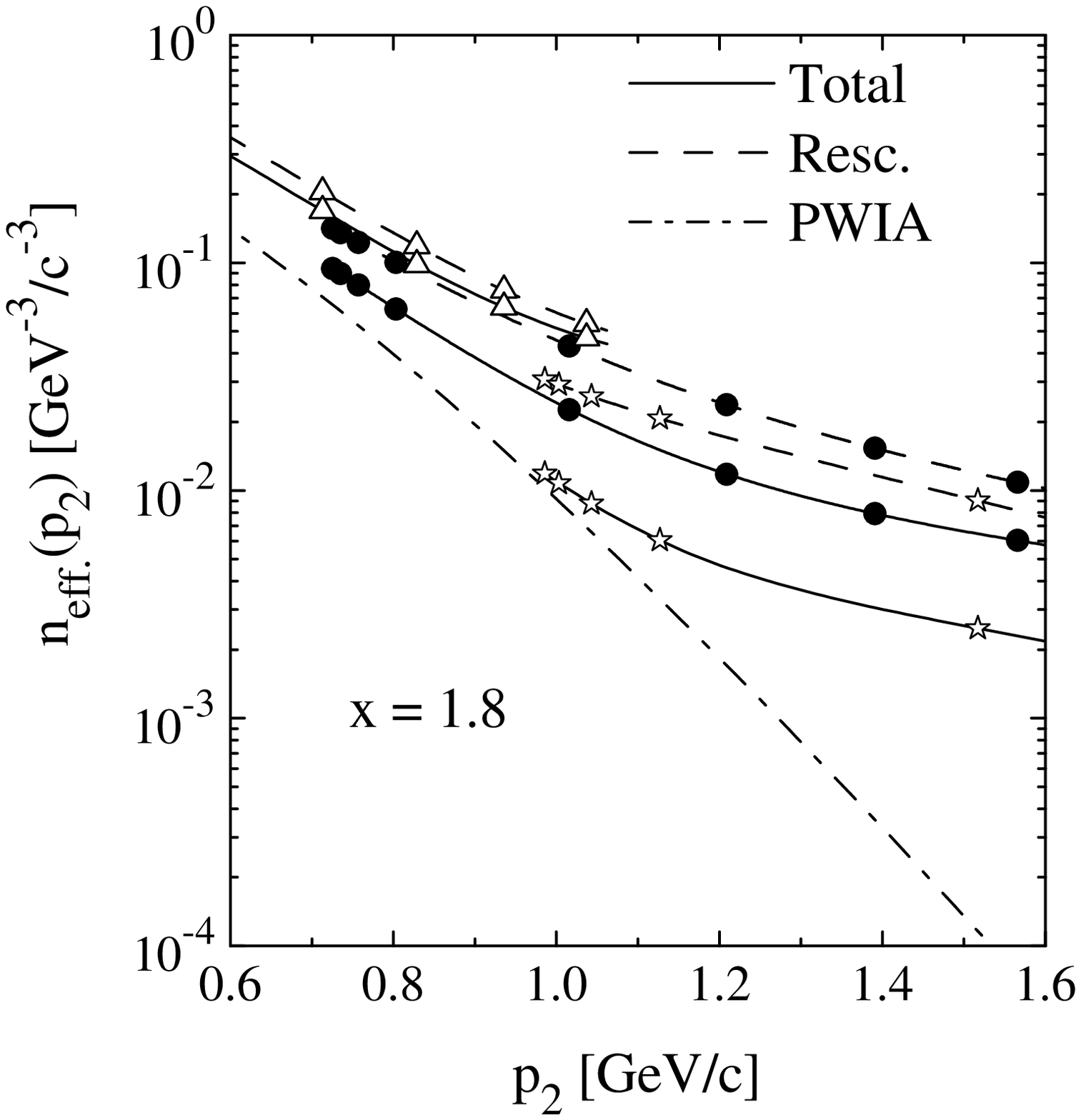}}
\caption{ The same as in Fig. \ref{Fig8} at $x=1.8$}
\label{Fig10}
\end{figure}

\begin{figure}[ht]
\epsfxsize 5.in
\centerline{\epsfbox{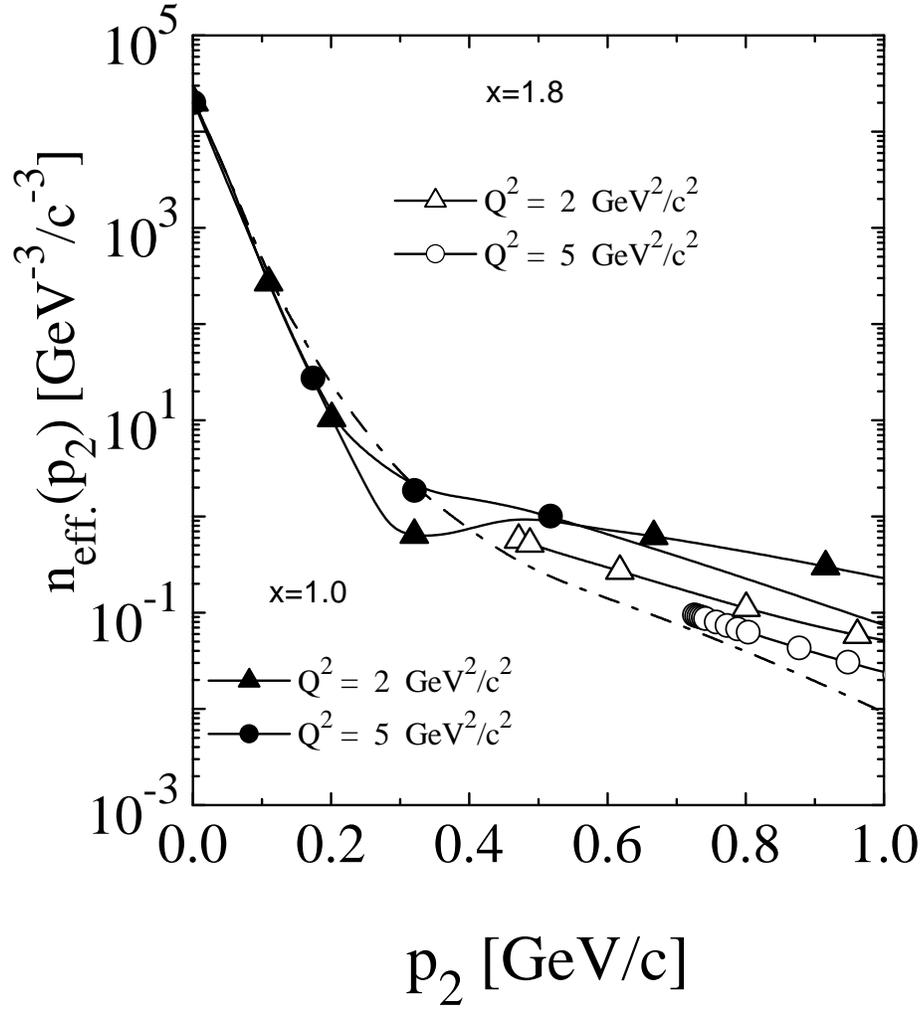}}
\caption{
The results of Figs. 7 and 9 are presented here together; the full symbols refer to
the effective momentum distribution
(Eq. \ref{ratio1}) calculated at $x=1$ and the open symbols to  $x=1.8$.
 The dot - dashed curve represents the  PWIA.}
\label{Fig11}
\end{figure}

\begin{figure}[ht]
\epsfxsize 6.in
\centerline{\epsfbox{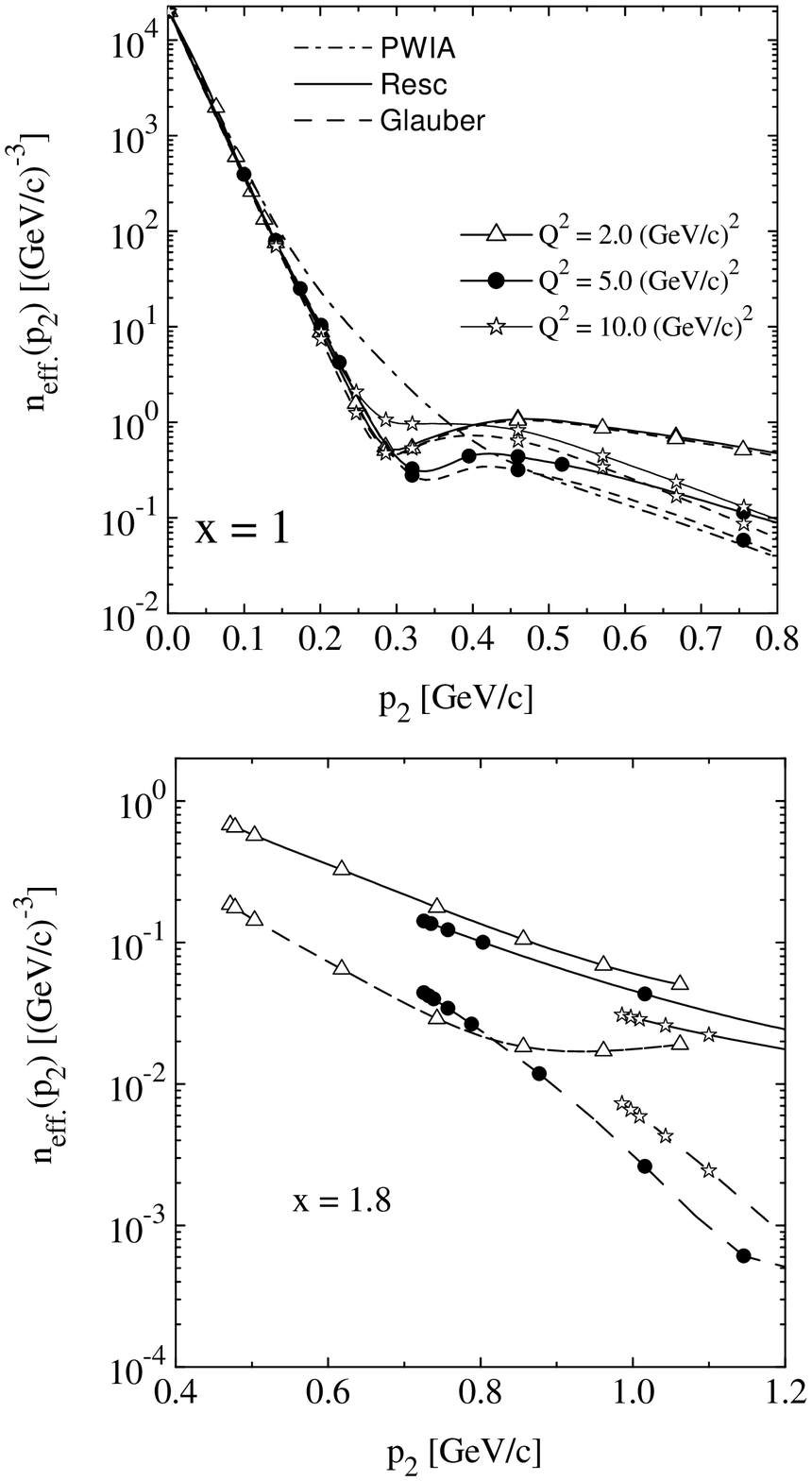}}
\caption{ Comparison between the effective momentum distributions (Eq. \ref{ratio1}) calculated within our approach
(full lines) and within the Glauber approximation (dashed lines); the dot-dashed line represents the
PWIA result}
\label{Fig12}
\end{figure}

\begin{figure}[ht]
\epsfxsize 4.in
\centerline{\epsfbox{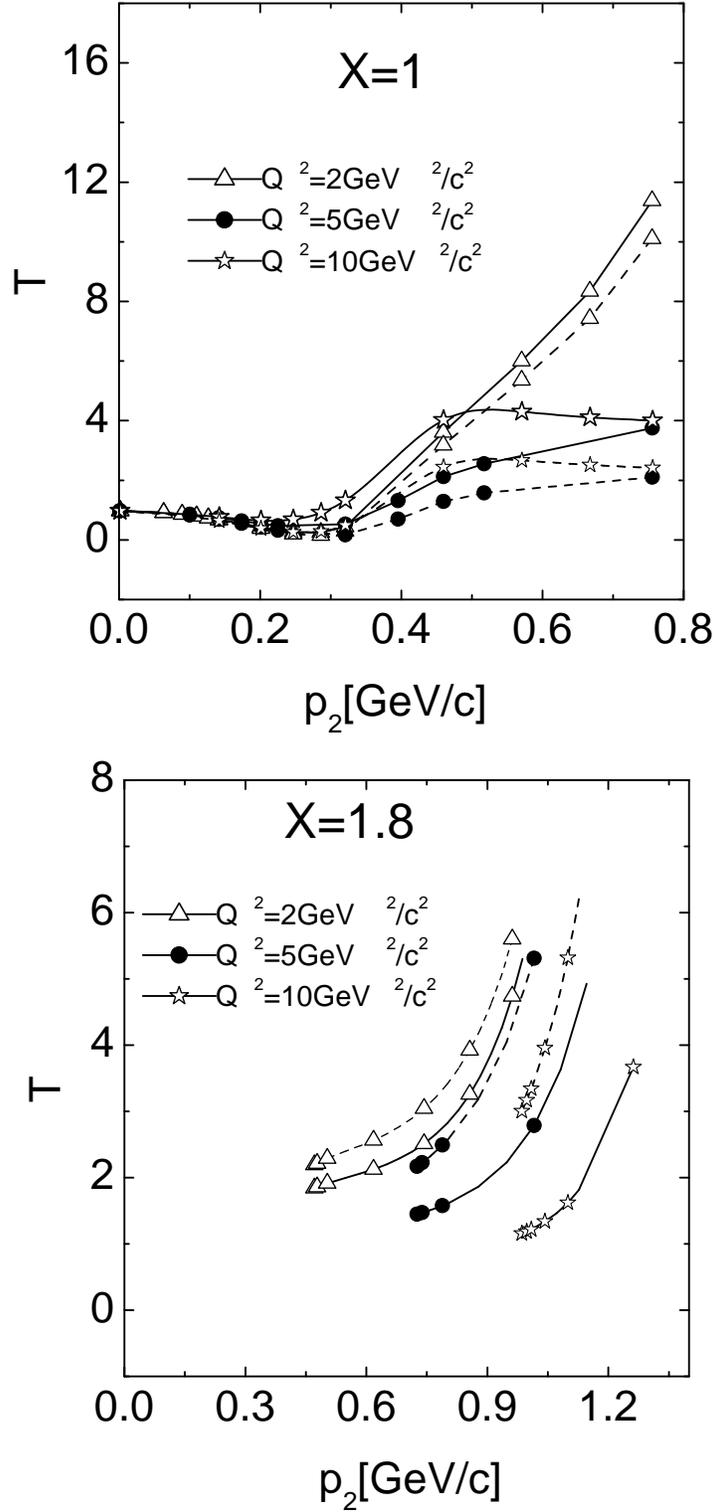}}
\caption{ The transparency T,  Eq. 
\ref{transp}, plotted  {\it vs.} $|{\bf p}_2| \equiv p_2$
  calculated at $x=1$ and $x=1.8$.
 The dashed
curves represent  the PWIA plus rescattering 
effects, whereas the full curves  include also
FFT effects.}
\label{Fig13}
\end{figure}

\end{document}